\documentclass[prl,twocolumn,showpacs,superscriptaddress,preprintnumbers,floatfix]{revtex4-1}

\usepackage{etex}
\usepackage{ifpdf}
\usepackage{hyperref}
\usepackage{dcolumn}
\usepackage{url}
\usepackage{amsmath}
\usepackage{amscd}
\usepackage{amsfonts}
\usepackage{amssymb}
\usepackage{bm}   
\usepackage{bbm}
\usepackage{verbatim}
\usepackage{stmaryrd}
\usepackage{amsthm}
\usepackage{xcolor}
\usepackage{setspace}

\usepackage{tikz}
\usetikzlibrary{arrows}
\usetikzlibrary{automata}
\usetikzlibrary{decorations.pathreplacing}
\usetikzlibrary{positioning}
\usetikzlibrary{plotmarks}
\usetikzlibrary{calc}
\usetikzlibrary{patterns}
\usetikzlibrary{external}
\usepackage{pgfplots}
\pgfplotsset{compat=newest}

\usepackage{graphicx}

\theoremstyle{plain}    
\theoremstyle{plain}    
\theoremstyle{plain}    
\theoremstyle{plain}    
\theoremstyle{plain}    
\theoremstyle{plain}    
\theoremstyle{plain}    
\theoremstyle{plain}    
\theoremstyle{plain}    
\theoremstyle{plain}    
\theoremstyle{plain}    
\theoremstyle{plain}

\newcommand{\eM}     {\mbox{$\epsilon$-machine}}
\newcommand{\eMs}    {\mbox{$\epsilon$-machines}}

\newcommand{\MeasSymbol}   { {X} }

\newcommand{\Past}      { \smash{\overleftarrow {\MeasSymbol}} }

\newcommand{\Future}    { \smash{\overrightarrow{\MeasSymbol}} }

\newcommand{\CausalState}       { \mathcal{S} }

\newcommand{\Prob}      {\Pr} 

\newcommand{\hmu}               {h_\mu}

\newcommand{\forward}{+}
\newcommand{\reverse}{-}
\newcommand{\forwardreverse}{\pm} 

\newcommand{\FutureCausalState} { {\CausalState}^{\forward} }

\newcommand{\PastCausalState}   { {\CausalState}^{\reverse} }

\newcommand{\lastindex}[2]{
  \edef\tempa{0}
  \edef\tempb{#2}
  \ifx\tempa\tempb
    \edef\tempc{#1}
  \else
    \edef\tempa{0}
    \edef\tempb{#1}
    \ifx\tempa\tempb
      \edef\tempc{#2}
    \else
      \edef\tempc{#1+#2}
    \fi
  \fi
  \tempc
}

\newcommand{\rhomu}{\rho_\mu}
\newcommand{\rmu}{r_\mu}
\newcommand{\bmu}{b_\mu}

\newcommand{\I}{\mathbf{I}}

\newcommand{\CSjoint}[1][,]{
   \edef\tempa{:}
   \edef\tempb{#1}
   \ifx\tempa\tempb
      \ensuremath{\FutureCausalState\!#1\PastCausalState}
   \else
      \ensuremath{\FutureCausalState#1\PastCausalState}
   \fi
}
\newcommand{\CSjointKL}[3][,]{
   \edef\tempa{:}
   \edef\tempb{#1}
   \ifx\tempa\tempb
      \ensuremath{\FutureCausalState_{#2}\!#1\PastCausalState_{#3}}
   \else
      \ensuremath{\FutureCausalState_{#2}#1\PastCausalState_{#3}}
   \fi
}

\newif\ifpm
\edef\tempa{\forwardreverse}
\edef\tempb{\pm}
\ifx\tempa\tempb
   \pmfalse
\else
   \pmtrue
\fi

\newcommand{\MeasSymbols}[2] {\MeasSymbol_{#1:#2}}

\renewcommand{\Past}{\MeasSymbols{}{0}}
\newcommand{\Present}{\MeasSymbol_0}
\renewcommand{\Future}{\MeasSymbols{1}{}}

\renewcommand{\H}{\operatorname{H}}
\renewcommand{\I}{\operatorname{I}}

\newcommand{\MS}{\MeasSymbol}

\begin{document}

\title{Chaos Forgets and Remembers:\\
Measuring Information Creation, Destruction, and Storage}

\author{Ryan G. James}
\email{rgjames@ucdavis.edu}
\affiliation{Complexity Sciences Center and Physics Department,
University of California at Davis, One Shields Avenue, Davis, CA 95616}

\author{Korana Burke}
\email{kburke@ucdavis.edu}
\affiliation{Complexity Sciences Center and Physics Department,
University of California at Davis, One Shields Avenue, Davis, CA 95616}

\author{James P. Crutchfield}
\email{chaos@ucdavis.edu}
\affiliation{Complexity Sciences Center and Physics Department,
University of California at Davis, One Shields Avenue, Davis, CA 95616}
\affiliation{Santa Fe Institute, 1399 Hyde Park Road, Santa Fe, NM 87501}

\date{\today}
\bibliographystyle{unsrt}

\begin{abstract}
The hallmark of deterministic chaos is that it creates information---the rate
being given by the Kolmogorov-Sinai metric entropy. Since its introduction half
a century ago, the metric entropy has been used as a unitary quantity to
measure a system's intrinsic unpredictability. Here, we show that it
naturally decomposes into two structurally meaningful components: A portion
of the created information---the ephemeral information---is forgotten and a
portion---the bound information---is remembered. The bound information is a
new kind of intrinsic computation that differs fundamentally from
information creation: it measures the rate of active information storage.
We show that it can be directly and accurately calculated via symbolic
dynamics, revealing a hitherto unknown richness in how dynamical systems
compute.

\vspace{0.1in}

\end{abstract}

\preprint{Santa Fe Institute Working Paper 13-09-030}
\preprint{arxiv.org:1309.5504 [nlin.CD]}

\maketitle
\

\def \r { 1.25cm }

\def \ipast { (-2.5, 0.25) -- +(2,0) arc (90:-90:\r) -- +(-2,0) arc
  (270:90:\r) }
\def \ifuture { (0.5, 0.25) -- +(2,0) arc (90:-90:\r) -- +(-2,0) arc
  (270:90:\r) }
\def \ipresent { (0, 0) circle (1.5cm) }
\def \ipresenta { (0, 0) circle (1.511cm) }

\colorlet {past_color}    {red}
\colorlet {pres_color}    {blue}
\colorlet {futu_color}    {black!30!green}

\colorlet {temp_color_1}  {red!50!blue}
\colorlet {temp_color_2}  {red!50!green}
\colorlet {temp_color_3}  {blue!50!green}

\colorlet {hmu_color}     {blue!67!green}
\colorlet {rhomu_color}   {temp_color_1!80!blue}
\colorlet {rmu_color}     {blue}
\colorlet {bmu_1_color}   {temp_color_1}
\colorlet {bmu_2_color}   {temp_color_3}
\colorlet {qmu_color}     {temp_color_1!67!green}
\colorlet {wmu_color}     {temp_color_2!57!blue}
\colorlet {sigmamu_color} {temp_color_2}

\def \opacity { 0.33 }

\pgfdeclarepatternformonly{bigcrosshatch}{\pgfqpoint{-1pt}{-1pt}}{\pgfqpoint{4pt}{4pt}}{\pgfqpoint{5pt}{5pt}}%
{
  \pgfsetlinewidth{0.4pt}
  \pgfpathmoveto{\pgfqpoint{5.1pt}{0pt}}
  \pgfpathlineto{\pgfqpoint{0pt}{5.1pt}}
  \pgfpathmoveto{\pgfqpoint{0pt}{0pt}}
  \pgfpathlineto{\pgfqpoint{5.1pt}{5.1pt}}
  \pgfusepath{stroke}
}

\setstretch{1.1}


The world is replete with systems that generate
information---information that is then encoded in a variety of ways:
Erratic ant behavior eventually leads to intricate, structured colony
nests \cite{Bona99a,Cama03a}; thermally fluctuating magnetic spins
form complex domain structures \cite{Binn92a}; music weaves theme,
form, and melody with surprise and innovation \cite{Simm96a}. We now
appreciate that the underlying dynamics in such systems is frequently
deterministic chaos \cite{strogatz1994nonlinear,Jose98a}. In others,
the underlying dynamics appears to be fundamentally stochastic
\cite{Stre09a}. For continuous-state systems, at least, one operational
distinction between deterministic chaos and stochasticity is found
in whether or not information generation diverges with measurement
resolution \cite{Gasp93a}. This result calls back to Kolmogorov's
original use \cite{Kolm59} of Shannon's mathematical theory of
communication \cite{Shan48a} to measure a system's rate of information
generation in terms of the metric entropy. Since that time, metric
entropy has been understood as a unitary quantity. Whether
deterministic or stochastic, it is a system's degree of
unpredictability. Here, we show that this is far too simple a
picture---one that obscures much.

To ground this claim, consider two systems. The first, a fair coin:
Each flip is independent of the others, leading to a simple
uncorrelated randomness. As a result, no statistical fluctuation is
predictively informative. For the second system consider a stock
traded via a financial market: While its price is unpredictable, the
direction and magnitude of fluctuations can hint at its future
behavior. (This, at least, is the guiding assumption of the now-global
financial engineering industry.) We make this distinction rigorous
here, dividing a system's information generation into a component that
is relevant to temporal structure and a component divorced from it. We
show that the temporal component captures the system's internal
information processing and, therefore, is of practical interest when
harnessing the chaotic nature of physical systems to build novel
machines and devices \cite{Ditto2010}. We first introduce the new
measures, describe how to interpret and calculate them, and then apply
them via a generating partition to analyze several dynamical
systems---the Logistic, Tent, and Lozi maps---revealing a previously
hidden form of active information storage.


We observe these systems via an optimal measuring instrument---called
a generating partition---that encodes all of their behaviors in a
\emph{stationary process}: A distribution $\Pr(\dots, \MS_{-2},
\MS_{-1}, \MS_0, \MS_1, \MS_2, \ldots)$ over a bi-infinite sequence of
random variables with shift-invariant statistics. A contiguous block
of observations $\MeasSymbols{t}{t+\ell}$ begins at index $t$ and
extends for length $\ell$. (The index is inclusive on the left and
exclusive on the right.) If an index is infinite, we leave it blank.
So, a process is compactly denoted $\Pr(\MS_:)$. Our analysis splits
$\MS_:$ into three segments: the \emph{present} $\Present$, a single
observation; the \emph{past} $\Past$, everything prior; and
\emph{future} $\Future$, everything that follows.

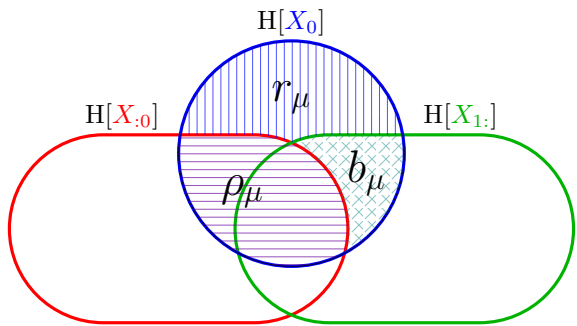
\begin{figure}
  \centering
  \begin{tikzpicture}[scale=1.00]
    \draw [very thick, past_color] \ipast;
    \draw [very thick, futu_color] \ifuture;
    \draw [very thick, pres_color] \ipresent;


    \begin{scope}
      \clip \ipast;
      \draw [pattern=horizontal lines, pattern color=rhomu_color, opacity=2*\opacity] \ipresent;
    \end{scope}
    \begin{scope}[even odd rule]
      \clip \ipast \ipresent;
      \clip \ifuture \ipresent;
      \draw [pattern=vertical lines, pattern color=rmu_color, opacity=2*\opacity] \ipresent;
    \end{scope}
    \begin{scope}[even odd rule]
      \clip \ifuture \ipast;
      \clip \ifuture;
      \draw [pattern=bigcrosshatch, pattern color=bmu_2_color, opacity=2*\opacity] \ipresent;
    \end{scope}

    \draw (-2.25, 0.5) node {{$\H[\textcolor{past_color}{\Past}]$}};
    \draw (2.25, 0.5) node {{$\H[\textcolor{futu_color}{\Future}]$}};
    \draw (0.0, 1.75) node {{$\H[\textcolor{pres_color}{\Present}]$}};

    \draw (0, 0.75) node {{\LARGE $\rmu$}};
    \draw (1.0, -0.25) node {{\LARGE $\bmu$}};
    \draw (-0.65, -0.5) node {{\LARGE $\rhomu$}};

  \end{tikzpicture}

  \caption{A process's I-diagram showing how the past $\Past$, present
    $\Present$, and future $\Future$ partition each other into seven
    distinct information \emph{atoms}. We focus only on the four
    regions contained in the \emph{present information} $\H[\Present]$
    (blue circle). That is, the present decomposes into three
    components: $\rhomu$ (horizontal lines), $\rmu$ (vertical lines),
    and $\bmu$ (diagonal crosshatching). The \emph{redundant}
    information $\rhomu$ overlaps with the \emph{past} $\H[\Past]$;
    the \emph{ephemeral} information $\rmu$ falls outside both the
    \emph{past} and the \emph{future} $\H[\Future]$. The \emph{bound}
    information $\bmu$ is that part of $\H[\Present]$ which is in the
    future yet not in the past. }
\label{fig:anatomy}
\end{figure}

The information-theoretic relationships between these three random
variable segments are graphically expressed in a Venn-like diagram,
known as an I-diagram~\cite{Yeun08a}; see Fig.~\ref{fig:anatomy}. The
rate $\hmu$ of information generation is the amount of new information
in an observation $\Present$ given all the prior observations $\Past$:
\begin{align}
  \hmu = \H[\Present | \Past]~,
\label{eq:entropy_rate}
\end{align}
where $\H[Y|Z]$ denotes the Shannon conditional entropy of random
variable $Y$ given variable $Z$. This quantity arises in various
contexts and goes by many names: e.g., the Shannon entropy rate and
the Kolmogorov-Sinai metric entropy, mentioned above~\cite{Gasp93a}.
The complement of the entropy rate is the \emph{predicted information}
$\rhomu$:
\begin{align}
  \rhomu = \I[\Past : \Present]~,
\label{eq:rhomu}
\end{align}
where $\I[Y:Z]$ denotes the mutual information between random
variables $Y$ and $Z$ \cite{Yeun08a}. Hence, $\rhomu$ is the
information in the present that can be predicted from prior
observations. Together, we have a decomposition of the information
contained in the present: $\H[\Present] = \hmu + \rhomu$.

A simple application of the entropy chain rule \cite{Yeun08a} to Eq.
(\ref{eq:entropy_rate}) leads us to a different view:
\begin{align}
  \hmu & = \I[\Present : \Future | \Past]
    + \H[\Present | \Past , \Future] \nonumber \\
  & = \bmu + \rmu ~.
\label{eq:anatomy}
\end{align}
This introduces two new information measures:
\begin{align}
\label{eq:bmu}
  \bmu &= \I[\Present : \Future | \Past] \text{~and~}\\
\label{eq:rmu}
  \rmu &= \H[\Present | \Past , \Future]~.
\end{align}
That is, created information ($\hmu$) decomposes into two parts:
information ($\bmu$) shared by the present and the future but not in
the past and information ($\rmu$) in the present but in neither the
past nor the future.

The $\rmu$ component was first studied by Verd\'u and
Weissman~\cite{Verdu2008} as the \emph{erasure entropy} (their $H^-$)
to measure information loss in erasure channels. To emphasize that it
is information existing only in a single moment---created and then
immediately forgotten---we refer to $\rmu$ as the \emph{ephemeral
  information}. The second component $\bmu$ we call the \emph{bound
  information} since it is information created in the present that the
system stores and that goes on to affect the future~\footnote{Our
  terminology avoids the misleading use of the phrase ``predictive
  information'' for $\bmu$. The latter is not the amount of
  information needed to predict the future. Rather, it is part of the
  \emph{predictable} information---that portion of the future which
  \emph{can} be predicted.}. It was first studied as a measure of
``interestingness'' in computational musicology by Abdallah and
Plumbley~\cite{Abdallah2009}. For a more complete analysis of this
decomposition, as well as computation methods and related measures,
see Ref.~\cite{James2011}.

Isolating the information $\H[\Present]$ contained in the present and
identifying its components provides the partitioning illustrated in
Fig.~\ref{fig:anatomy}. This is a particularly intuitive way of
thinking about the information contained in an observation. While,
some behavior ($\rhomu$) can be predicted, the rest ($\hmu = \bmu +
\rmu$) cannot. Of that which cannot be predicted, some ($\bmu$) plays
a role in the future behavior and some ($\rmu$) does not. As such,
this is a natural decomposition of a time series; one that results in
a semantic dissection of the entropy rate.

By way of an example, consider a few simple processes and how their
present information decomposes into these three components. A periodic
process of alternating $0$s and $1$s ($\ldots01010101\ldots$) has
$\H[\Present] = 1$ bit since $0$s and $1$s occur equally often. Given
a prior observation, one can accurately predict exactly which symbol
will occur next and so $\H[\Present] = \rhomu = 1$ bit, while $\rmu =
\bmu = 0$ bits. On the other extreme is a fair coin flip. Again, each
outcome is equally likely and so $\H[\Present] = 1$ bit. However, each
flip is independent of all others and so $\H[\Present] = \rmu = 1$
bit, while $\rhomu = \bmu = 0$ bits.

Between these two extrema lie interesting processes: those with
\emph{stochastic structure}. Processes expressing a fixed template,
like the periodic process above, contain a finite amount of
information. Those with stochastic structure, however, constantly
generate information and store it in the form of patterns. Being
neither purely predictable nor independently random, these patterns
are captured by $\bmu$. The more intricate the organization, the
larger $\bmu$. More to the point, generating these patterns requires
intrinsic computation in a system---information creation, storage, and
transformation \cite{Crut88a}. We propose $\bmu$ as a simple method of
discovering this type of physical computation: Where there are
intricate patterns, there is sophisticated processing.


How useful is the proposed decomposition and its measures? To answer
this we analyze several discrete-time chaotic dynamical systems---the
Logistic and Tent maps of the interval and the Lozi map of the
plane---uncovering a number of novel properties embedded in these
familiar and oft-studied systems. As an independent calibration for
the measures, we employ Pesin's theorem \cite{Pesin1977}: $\hmu$ is
the sum of the positive Lyapunov characteristic exponents (LCEs). The
maps here have at most one positive LCE $\lambda$, so $\hmu= \max
\{0,\lambda\}$. The symbols $s_0, s_1, s_2, \ldots, s_N$ for each
process we analyze come from a generating partition. We produce a long
sample of $N \approx 10^{10}$ symbols, extracting subsequence
statistics via a sliding window \footnote{Window width is adaptively
  chosen in inverse proportion to the LCE. When the latter is low we
  use a longer window than when the system is fully chaotic. The
  minimum window width of $L = 31$ and adaptive widths were chosen so
  that numerical estimates varied by less than $0.01\%$ when the width
  is incremented.}. Each window consists of a past, present, and
future symbol sequence and we estimate $\rmu$ and $\bmu$ using
truncated forms of Eqs. (\ref{eq:bmu}) and (\ref{eq:rmu}).

\begin{figure}
  \centering
  \includegraphics[width=\columnwidth]{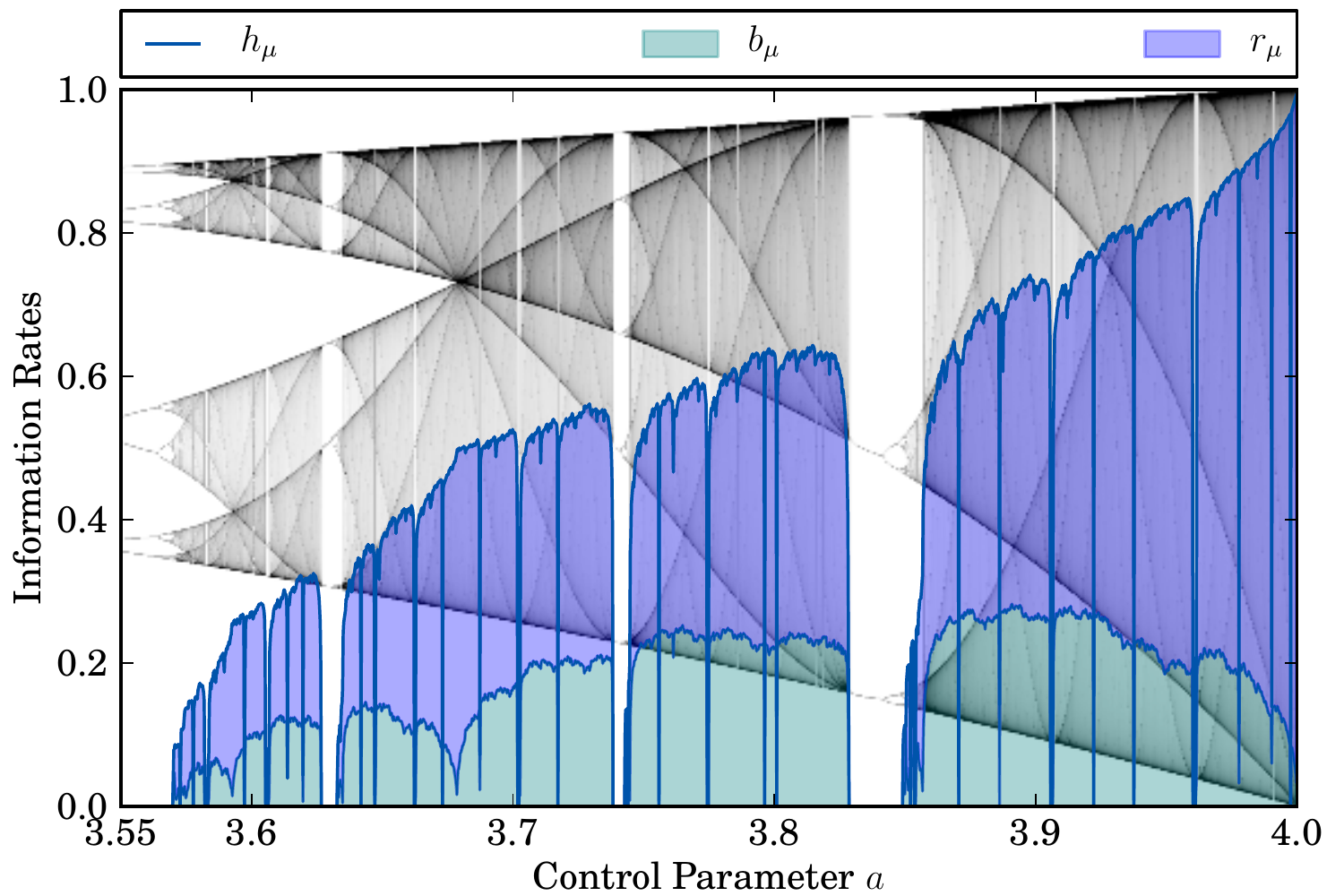}
  \caption{Logistic map information anatomy as a function of control
    parameter $a$: Bound information $\bmu$ is the lower (green
    shaded) component; ephemeral information $\rmu$ is the upper (blue
    shaded) component. Entropy rate is the top (blue) line: $\hmu =
    \bmu + \rmu$. As reference to the dynamical behavior, the map's
    bifurcation diagram is displayed in the background. }
\label{fig:logistic}
\end{figure}

Consider first the Logistic map, perhaps one of the most studied
chaotic systems:
\begin{align}
  x_{n+1} = a x_n (1 - x_n)
  ~,
\label{eq:logistic}
\end{align}
where $a \in [0,4]$ is the control parameter and the initial condition
is $x_0 \in [0,1]$. Its generating partition is defined by:
\begin{align}
  \label{eq:OneDMapPartition}
  s_n = \begin{cases}
          0 & \mbox{if } x_n < \frac{1}{2} \\
          1 & \mbox{if } x_n \ge \frac{1}{2}
        \end{cases}
  ~.
\end{align}
Figure~\ref{fig:logistic} shows the resulting measures as a function
of control $a$, with the map's bifurcation diagram displayed in the
background for reference.

The first point of interest is that the system's information
generation is, in fact, a mixture of ephemeral ($\rmu$) and bound
($\bmu$) informations at nearly all chaotic ($\hmu > 0$) parameter
values. The second is that the division into the two components varies
in a nontrivial way as a function of the control parameter $a$.
Moreover, the boundary between the two appears nondifferentiable. At
first blush, this is not surprising given that their sum
$\hmu$~($=\lambda$) is known to be nondifferentiable. Finally, $\bmu$
vanishes nontrivially only at parameters that coincide with the merging of the
chaotic bands (e.g., $a = 4.0, 3.67857\ldots, 3.59257\ldots, \ldots$).
Thus, the information generated by the Logistic map at these
parameters is entirely forgotten.

\begin{figure}
  \centering
  \includegraphics[width=\columnwidth]{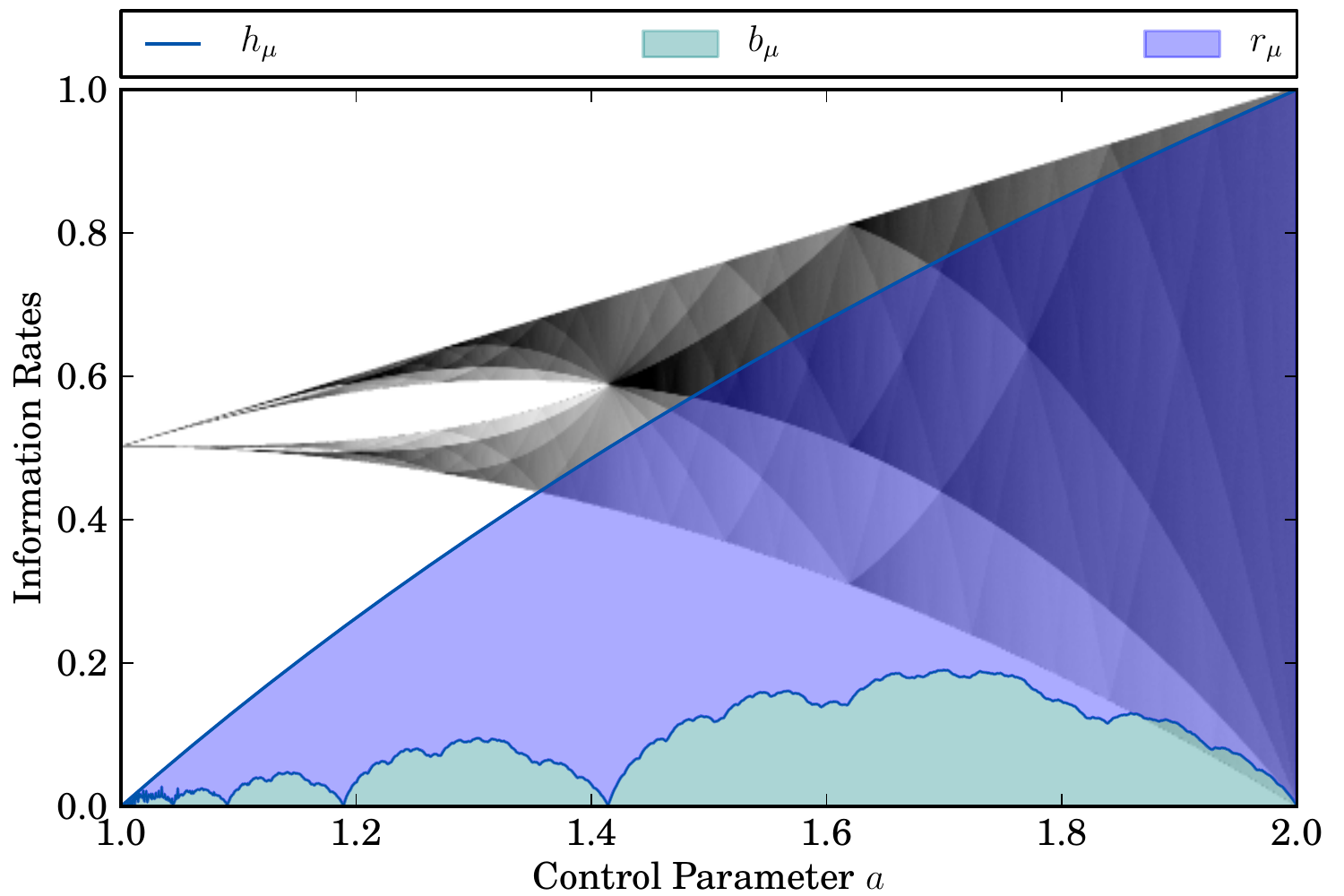}
  \caption{Tent map information anatomy: Although $\hmu = \bmu + \rmu$
    is a smooth function of control---$\hmu = \log_2 a$---the
    decomposition into bound and ephemeral informations is not.
    Graphics layout as in previous figure. }
\label{fig:tent}
\end{figure}

Is the complex and nondifferentiable boundary between $\rmu$ and
$\bmu$ simply a consequence of the entropy rate's complicated behavior
or due a dynamical mechanism distinct from information creation? We
answer this by analyzing the Tent map:
\begin{align}
  \label{eq:tent}
  x_{n+1} = \frac{a}{2} \left( 1 - 2 \left| x_n - \frac{1}{2} \right| \right)
  ~,
\end{align}
where $a \in [0,2]$ is the control parameter. The generating partition for the
Tent map is the same as for the Logistic map. Since the Tent map is piecewise
linear, its Lyapunov exponent is simply $\lambda = \log_2{a}$ and, by Pesin's
theorem, so is the information generation $\hmu = \log_2{a}$; a rather smooth
parameter dependence. As a result, the intricate structures exhibited in the
Tent map's bifurcation diagram cannot be resolved by studying solely the
behavior of the Lyapunov exponent (or $\hmu$) itself.
Figure~\ref{fig:tent} demonstrates that, despite the entropy rate's simple
logarithmic dependence on control, its decomposition $\hmu = \bmu+ \rmu$ is not
a smooth function of $a$. To emphasize, in sharp contrast with $\hmu$'s
simplicity, $\rmu$ and $\bmu$ again appear nondifferentiable---a complexity
masked by the smooth $\hmu$. Thus, the two informational components capture a
property in the chaotic system's behavior that is both quantitatively and
qualitatively new. As with the Logistic map, we once again find that the bound
information vanishes and that all of the information the Tent map generates is
forgotten ($\hmu = \rmu$) at parameters corresponding to merging of chaotic
bands ($a = 2^{1/2^k}, k = 0, 1, 2, \ldots$). In the Supplementary Materials we
show how to calculate $\bmu$ and $\rmu$ in closed form for the Tent map at
Misiurewicz parameters.

\begin{figure}
  \centering
  \includegraphics[width=\columnwidth]{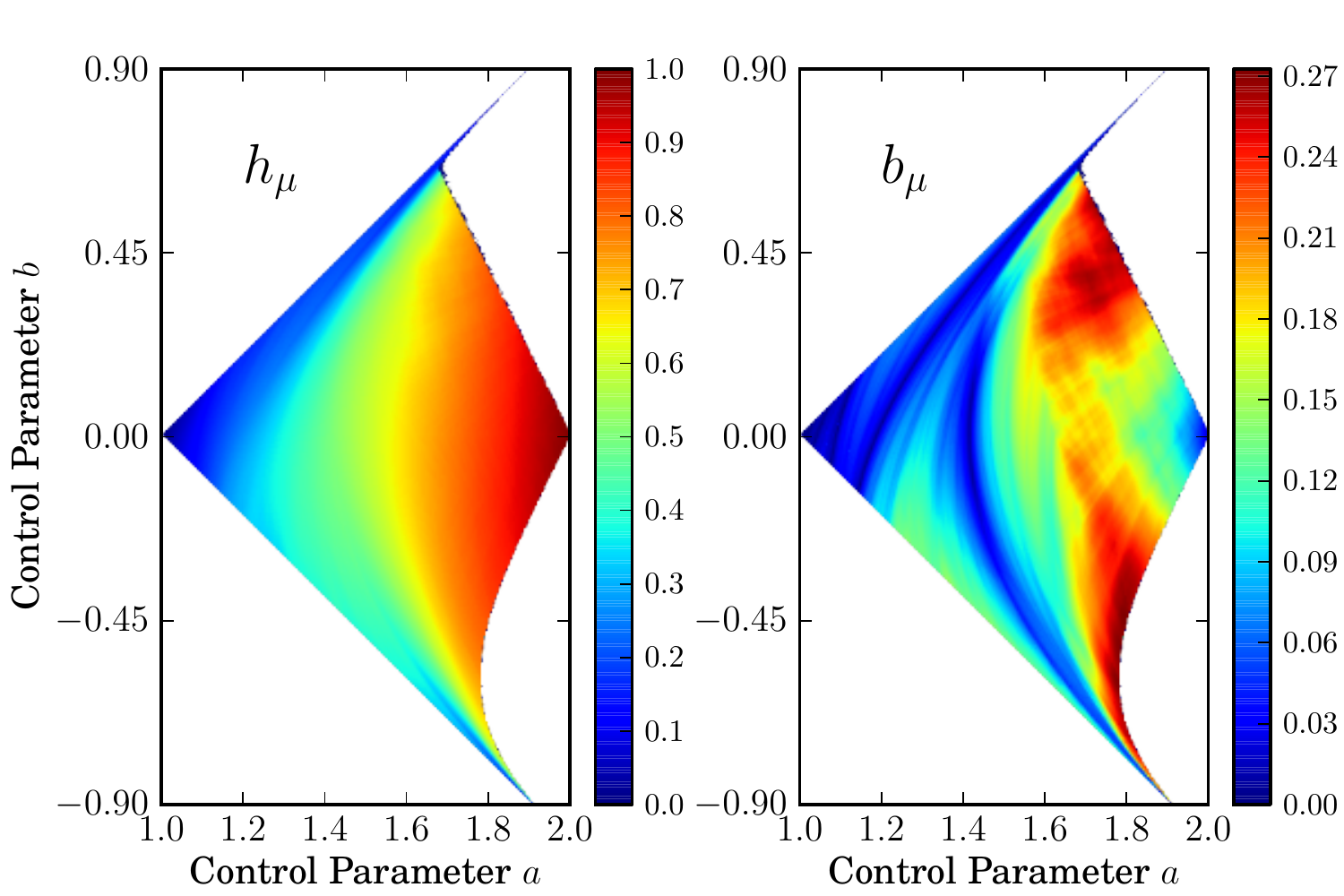}
  \caption{Lozi map information anatomy: (Left) $\hmu$ as a function
    of controls $a$ and $b$. (Right) $\bmu$ similarly. $\bmu$ is
    maximized on the upper-right and lower-right edges of the $a$-$b$
    region that supports an attractor near the origin. }
\label{fig:lozi}
\end{figure}

To explore how these measures apply more generally, we extend
information anatomy to two dimensions by analyzing the Lozi map:
\begin{align}
  \label{eq:lozi}
  x_{n+1} &= 1 - a \left| x_n \right| + y_n \\
  y_{n+1} &= b x_n~. \nonumber
\end{align}
The map exhibits an attractor near the origin within a diamond-shaped
parameter region inside $(a,b) \in [1,2] \times [-0.9,0.9]$. Note that
when $b=0$ the map becomes isomorphic to the Tent map. The generating
partition is given by:
\begin{align}
  \label{eq:LoziPartition}
  s_n = \begin{cases}
          0 & \mbox{if } x_n < 0 \\
          1 & \mbox{if } x_n \ge 0
        \end{cases}
  ~.
\end{align}

Figure~\ref{fig:lozi} shows $\hmu$ (left) and $\bmu$ (right) in the
attracting parameter region. Mirroring the Tent map, the Lozi map's
entropy rate varies smoothly over the attractor region, whereas $\bmu$
varies in a more complicated manner. There are swaths of low $\bmu$
corresponding to ``fuzzy'' mergings of chaotic bands. Notably, while
the maximal $\hmu$ occurs along the line $b=0$, maximal $\bmu$ occurs
far from $b=0$. Hence, large $\hmu$ does not necessarily imply large
bound information $\bmu$.


To sum up, we showed that a process's information creation rate
decomposes, via a chain rule, into two structurally meaningful
components. The components, the ephemeral information $\rmu$ and the
bound information $\bmu$, provide direct insights into a system's
behavior without detailed modeling or appealing to domain-specific
knowledge. That is to say, they are relatively easily defined measures
that can be straightforwardly estimated. More to the point, however,
$\bmu$ is a strong indicator of intrinsic computation. While related
to information generation, we demonstrated that it captures a
different kind of informational processing---a mechanism that actively
stores information.

Concretely, decomposing information creation in the symbolic dynamics
of the Logistic, Tent, and Lozi systems delineated the topography of their
intrinsic-computation landscape. Awareness of this rich (and previously
hidden) landscape will lead to improved engineering of natural systems as
substrates for information processing \cite{Ditto2010}. And, it will lead
to an expanded understanding of evolved information processing systems, such
as the linguistic processes comprising human natural languages. A sequel will
develop the decomposition further, including a geometric interpretation of
active information storage that parallels the geometric view of information
creation expressed in the Lyapunov exponents.


KB is supported by a UC Davis Chancellor's Post-Doctoral Fellowship.
This work was partially supported by ARO grant W911NF-12-1-0288.

\bibliography{chaos_anatomy}

\cleardoublepage

\appendix

\begin{center}
{\emph{Chaos Forgets and Remembers:\\
Measuring Information Creation, Destruction, and Storage}\\
\bf Supplementary Material}\\
Ryan G. James, Korana Burke, and James P. Crutchfield
\end{center}

\section*{Computing Bound and Ephemeral Informations Analytically}
\label{sec:comp-bmu-analyt}

\begin{figure}[b]
  \centering
  \includegraphics[width=0.95\columnwidth]{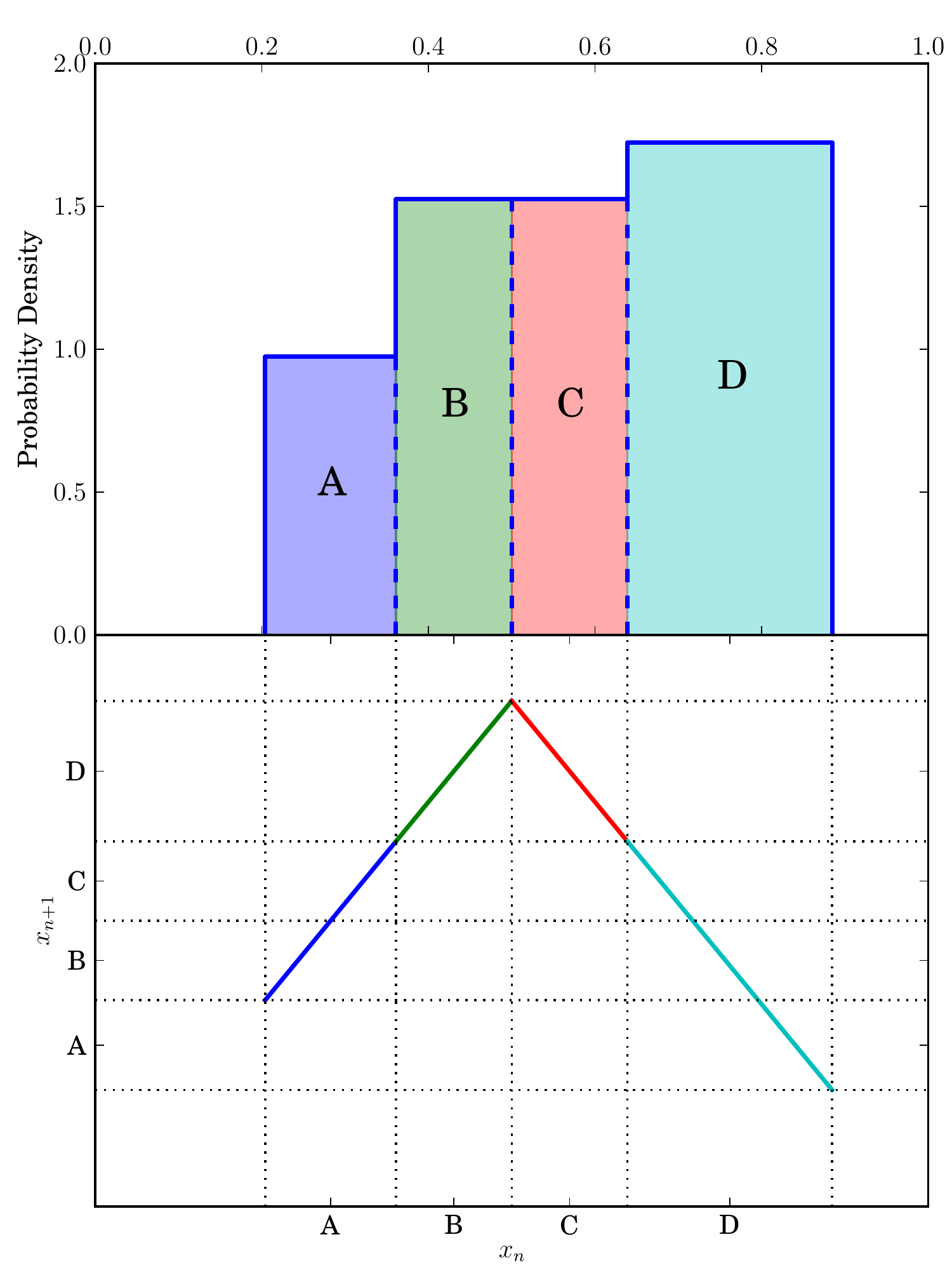}
  \caption{(Above) Tent map's invariant distribution at parameter $a$
    as defined in Eq.~(\ref{eq:misiurewicz}), consisting of three
    contiguous, uniformly distributed parts, each differently colored
    for clarity. (Below) The same colors superimposed on the map
    itself along with (dotted line) guides to show that the uniform
    components of the invariant distribution do indeed form a Markov
    partition. }
\label{fig:markov_partition}
\end{figure}

To obviate the data requirements for accurately estimating $\bmu$ from a time
series, we present a method for computing it analytically, in
closed form. An analytic expression is possible if one can construct
forward-time and reverse-time models of the system. These models are
sufficient statistics of the past about the present and future, and
the future about the present and past, respectively. Here, we use
\eMs\ [S1, S2], which are the minimal sufficient statistics. From
these, $\bmu$ can be computed via:
\begin{align}
  \label{eq:analytic}
  \bmu = \I[\Present : \CausalState^+_0 | \CausalState^-_1]
  ~,
\end{align}
where $\CausalState^+_0$ is the forward \eM's state random variable at
time $0$---the minimal sufficient statistic of the past about the
present and future---and $\CausalState^-_1$ is the reverse-time \eM's
state random variable at time $1$---the minimal sufficient statistic
of the future about the present and future. Due to their standing as
sufficient statistics, these states stand in for the future $\Future$
and the past $\Past$ of Eq. (\ref{eq:bmu}).

We explicitly implement this calculation for one parameter value
of the Tent map. In particular, consider the Misiurewicz point where
$f^4(\tfrac{1}{2}) = f^5(\tfrac{1}{2})$. Solving this constraint gives
parameter value:
\begin{align}
  \label{eq:misiurewicz}
  a &= \alpha + \frac{2}{3 \alpha} \\
    &= 1.76929235\ldots \nonumber
\end{align}
where $\alpha = \sqrt[3]{\sqrt{\frac{19}{27}}+1}$. There, the Tent map
admits a Markov partition [S3], as Fig.~\ref{fig:markov_partition}
demonstrates. From this, a Markov chain is constructed and the
generating partition overlaid. The result is the hidden Markov model
of Fig.~\ref{fig:markov_chain}(right) that exactly describes the
map's symbolic dynamics stochastic process.

\begin{figure}
  \centering
  \includegraphics[width=0.45\columnwidth]{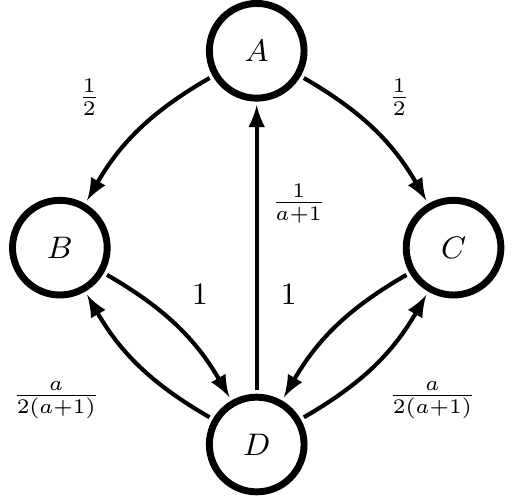}%
  \includegraphics[width=0.45\columnwidth]{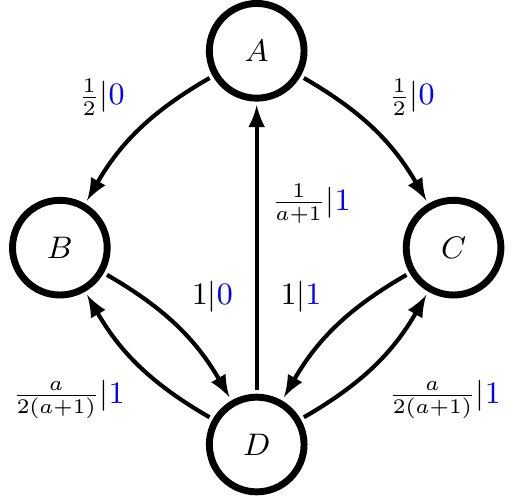}
  \caption{(Left) Markov chain induced by the Markov partition.
    (Right) The generating partition applied to the transitions,
    resulting in a hidden Markov model that describes the Tent map's
    symbolic stochastic process. }
\label{fig:markov_chain}
\end{figure}

Equation~(\ref{eq:analytic}) requires the model to be a sufficient
statistic to calculate $\bmu$. And so, we transform the hidden Markov
model of Fig.~\ref{fig:markov_chain}(right) to one that is
\emph{unifilar} and, in particular, to the \emph{\eM} of
Fig.~\ref{fig:em}.

\begin{figure}
  \centering
  \includegraphics[width=0.9\columnwidth]{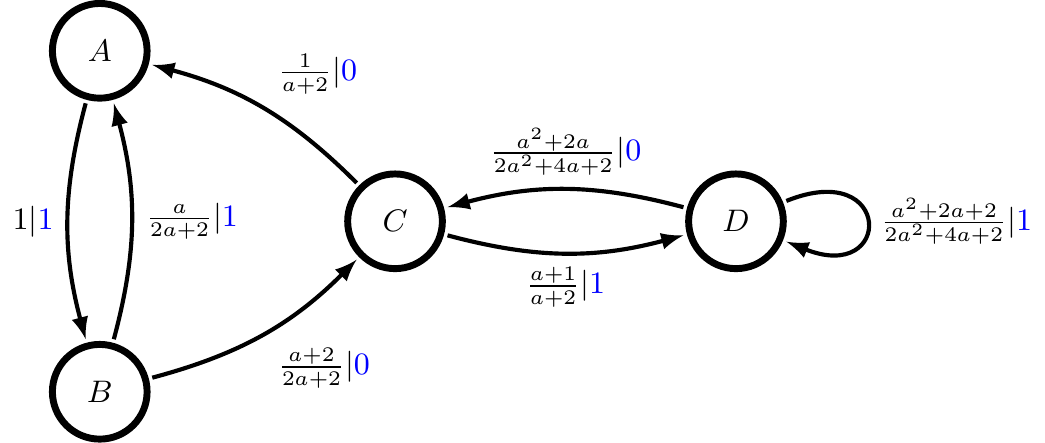}
  \caption{The (unifilar) \eM\ for the hidden Markov model of
    Fig.~\ref{fig:markov_chain}(right). That is, for each state there
    is at most a single outgoing transition labeled with each symbol.
    This makes the states a function of the past $\Past$ and so allows
    for the required calculation. }
\label{fig:em}
\end{figure}

As the final step, we construct the process's \emph{bidirectional machine}
[S1, S2] from the \eM; the result is shown in
Fig.~\ref{fig:bidirectional}. Then, from it we calculate the joint
distribution $\Prob(\CausalState^+_0, \CausalState^-_0, \Present,
\CausalState^+_1, \CausalState^-_1)$. This, in turn, allows one to
calculate $\bmu = \H[\Present | \CausalState^+_0 , \CausalState^-_1]$
and $\rmu = \I[\Present : \CausalState^-_1 | \CausalState^+_0]$. We
find that the Tent map at the Misiurewicz parameter $a$ has the
following information measures (in bits per step):
\begin{align*}
  \label{eq:values}
\hmu &= \log_2{a} = \log_2 \left(\frac{\sqrt[3]{9+\sqrt{57}} +
        \sqrt[3]{9-\sqrt{57}}}{3^{\frac{2}{3}}}\right) \\
  &= 0.823172\ldots \\
\rmu &= \frac{1}{4} \left( 3 - \frac{2}{a+1} - \frac{4}{a + 2}
  + \frac{9}{2a + 3} \right) \\
  &= \frac{1}{9} \left(\frac{\sqrt[3]{207
  \sqrt{57}-1349}}{19^{2/3}}-\frac{32}{\sqrt[3]{19 \left(207
  \sqrt{57}-1349\right)}}+7\right) \\
  &= 0.648258\ldots \\
\bmu &= \hmu - \rmu \\
     &= 0.174915 \ldots
  ~.
\end{align*}

\begin{figure}
  \centering
  \includegraphics[width=0.9\columnwidth]{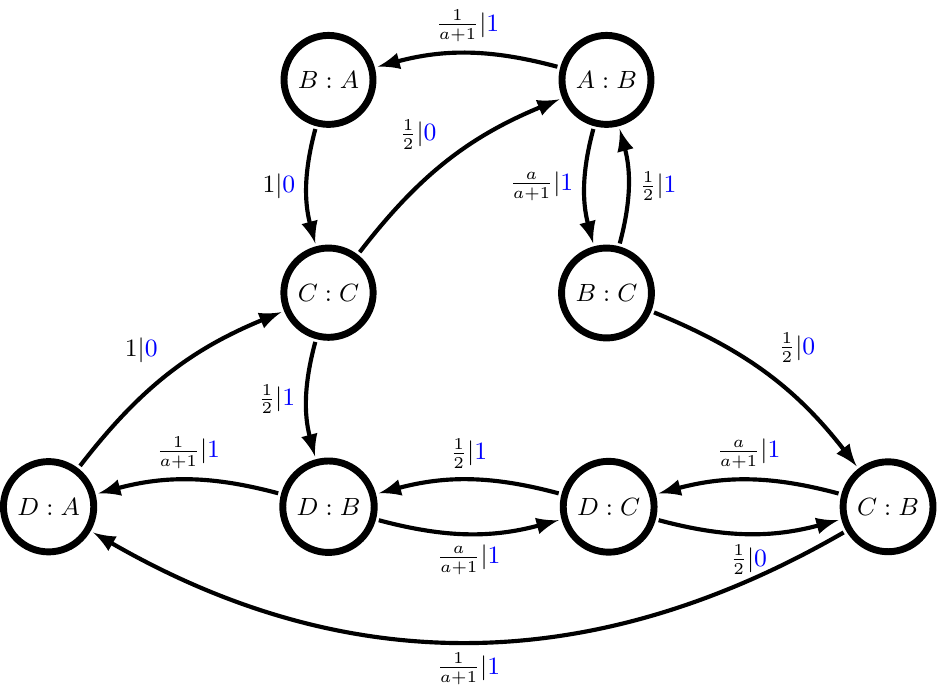}
  \caption{Bidirectional machine of the stochastic process generated
    by the Tent map's symbolic dynamics at the Misiurewicz parameter
    $a$. Its states are pairs $\CausalState^+_0 : \CausalState^-_0$.
    By utilizing the dynamic (the edges that connect states), it
    allows one to directly calculate $\H[\Present | \CausalState^+_0,
    \CausalState^-_1]$ and, thus, $\rmu$ and $\bmu$. }
\label{fig:bidirectional}
\end{figure}

\begin{center}
Supplementary References
\end{center}
S1. J. P. Crutchfield, C. J. Ellison, and J. R. Mahoney, ``Time's
Barbed Arrow: Irreversibility, Crypticity, and Stored Information'',
Phys. Rev. Lett. {\bf 103}:9 (2009) 094101.

\noindent
S2. C. J. Ellison, J. R. Mahoney, and J. P. Crutchfield, ``Prediction,
Retrodiction, and the Amount of Information Stored in the Present'',
J. Stat. Phys. {\bf 136}:6 (2009) 1005--1034.

\noindent
S3. D. Lind and B. Marcus. \emph{An Introduction to Symbolic
  Dynamics and Coding}. Cambridge University Press, 1999.

\end{document}